\documentclass[conference]{IEEEtran}
\IEEEoverridecommandlockouts
\usepackage{cite}
\usepackage{algorithm2e}
\usepackage{algorithmic}
\usepackage{graphicx}
\usepackage{array, makecell} 
\usepackage{diagbox}
\usepackage{subcaption,multicol}
\usepackage{multirow}

\usepackage{epstopdf}
\usepackage{epsfig}
\usepackage{float}
\usepackage{enumitem}
\usepackage{amssymb,amsmath,amsfonts,bm,epsfig,graphicx,theorem}
\usepackage[dvipsnames]{xcolor}
\newcommand{\comment}[1]{}

\usepackage{comment}

\newcommand{\mv}{\mathbf{m}}

\newcommand{\sv}{\mathbf{s}}
\newcommand{\xv}{\mathbf{x}}

\newcommand{\zv}{\mathbf{z}}

\newcommand{\Dc}{\mathcal{C}}

\newcommand{\argmin}{\mathop{\mbox{\rm arg\,min}}}

\newcommand{\ssf}[1]{\textrm{$\sf{#1}$}{}}

\newcommand{\vect}[1]{\mathbf{#1}}

\newcommand{\norm}[1]{\left \| #1 \right \|}

\newcommand{\Expect}{\mathbb{E}}

\newcommand{\pth}[1]{\left( #1 \right)}

\newcommand{\sth}[1]{\left\{ #1 \right\}}

\ifodd 1
\newcommand{\congr}[1]{{\color{red}#1}}
\else
\newcommand{\congr}[1]{#}
\fi

\ifodd 1
\newcommand{\congc}[1]{{\color{red}(Cong: #1)}}
\else
\newcommand{\congc}[1]{}
\fi

\newcommand{\mypara}[1]{{\smallskip \noindent \bf #1}\hspace{0.1in}}

\def\BibTeX{{\rm B\kern-.05em{\sc i\kern-.025em b}\kern-.08em
    T\kern-.1667em\lower.7ex\hbox{E}\kern-.125emX}}

\begin{document}

\title{An Autoencoder-Based Constellation Design for AirComp in Wireless Federated Learning
\thanks{The work is partially support by the US National Science Foundation under award CNS-2002902, CPS-2313110, ECCS-2143559, ECCS-2033671, SII-2132700, and the Commonwealth Cyber Initiative (CCI) of Virginia under Award VV-1Q23-005.}
%
 }

\author{\IEEEauthorblockN{Yujia~Mu, Xizixiang~Wei,  Cong~Shen}
\IEEEauthorblockA{Charles L. Brown Department of Electrical and Computer Engineering, University of Virginia
}
}

\maketitle

\begin{abstract}
Wireless federated learning (FL) relies on efficient uplink communications to aggregate model updates across distributed edge devices. Over-the-air computation (a.k.a. AirComp) has emerged as a promising approach for addressing the scalability challenge of FL over wireless links with limited communication resources. Unlike conventional methods, AirComp allows multiple edge devices to transmit uplink signals simultaneously, enabling the parameter server to directly decode the average global model. However, existing AirComp solutions are intrinsically analog, while modern wireless systems predominantly adopt digital modulations. Consequently, careful constellation designs are necessary to accurately decode the sum model updates without ambiguity. In this paper, we propose an end-to-end communication system supporting AirComp with digital modulation, aiming to overcome the challenges associated with accurate decoding of the sum signal with constellation designs. We leverage autoencoder network structures and explore the joint optimization of transmitter and receiver components. Our approach fills an important gap in the context of accurately decoding the sum signal in digital modulation-based AirComp, which can advance the deployment of FL in contemporary wireless systems.
\end{abstract}

\begin{IEEEkeywords}
Federated Learning; Constellation Design; Autoencoder.
\end{IEEEkeywords}

\section{Introduction}
\label{sec:intro}
In wireless federated learning (FL) \cite{mcmahan2017fl,wei2022tccn,zheng2020design}, efficient communication plays a crucial role in facilitating model aggregation across a large number of distributed edge devices that have limited communication resources. The model aggregation process involves computing the (weighted) average of local model updates and transmitting them from selected edge devices to the parameter server (PS). To address the scalability challenge of FL over wireless communications, one promising approach is \emph{over-the-air computation}, also known as \emph{AirComp} \cite{zhu2019broadband,yang2020federated,amiri2020machine}. Unlike the conventional approach of decoding individual local models from each edge device and then aggregating them, AirComp enables multiple edge devices to transmit the uplink signals simultaneously, in a superimposed manner, hence allowing the FL server to directly decode the average global model.

By employing AirComp, edge devices can simultaneously transmit model parameters, thereby significantly reducing the uplink communication cost regardless of the number of participating edge devices. However, it is important to note that the prevalent mode of communication in modern wireless systems is through digital modulation, whereas existing AirComp systems are analog in nature. This poses challenges in deploying AirComp techniques in contemporary wireless systems that predominantly employ digital modulations. Consequently, careful consideration must be given to modulation design to accurately decode the sum via the constellation design.  It is crucial to avoid a scenario where a specific point in the sum constellation corresponds to multiple different sums of the individual constellations. However, achieving this objective is a non-trivial task, as existing constellation designs have not considered decoding the sum.

Autoencoder (AE) \cite{mcclelland1987parallel} network structures exhibit similarities to the mathematical models of communication, as their encoder and decoder networks naturally resemble key components of the transmitter and receiver pairs in a communication system. Previous research has explored the concept of treating the communication system as an autoencoder to jointly optimize the transceiver \cite{o2017introduction}, while others have investigated the use of autoencoders for constellation design \cite{zheng2020design, omidi2021geometric, lin2022machine}. However, it is worth noting that none of these prior studies specifically focuses on the challenge of accurately decoding the sum. Given this limitation, we propose a novel end-to-end communication system that leverages AirComp for digital transmissions in the uplink wireless FL phase. The autoencoder-based approach allows us to overcome the existing challenges associated with the constellation design, particularly in accurately decoding the sum signal from individual constellations.

The remainder of this paper is organized as follows. 
The system model that captures the fading effects in wireless FL is described in Section~\ref{sec:sys_model}. The proposed end-to-end communication system and its autoencoder design for high-order modulations are presented in Section~\ref{sec:end_to_end}. Experimental results are given in Section~\ref{sec:sim}, followed by the conclusions in Section~\ref{sec:conc}.




\section{System Model}
\label{sec:sys_model}
We start by introducing the fundamental optimization problem in ML and proceed to explain the standard pipeline for distributed model training, with a particular emphasis on the celebrated \textsl{FedAvg} algorithm. Following this, we analyze the uplink communication model and discuss the influence of channel fading effects and receiver processing. It is crucial to understand these aspects as they directly impact the performance of wireless communication systems. Moreover, we highlight the limitations and challenges associated with existing constellations, which motivates the need for our work. 

\subsection{The Distributed SGD Problem}
\label{sec:modelFL}

In our research, we focus on studying the standard empirical risk minimization (ERM) problem in machine learning (ML):
\begin{equation} \label{eqn:erm}
\min_{\xv \in \mathbb{R}^d} F(\xv) = \min_{\xv \in \mathbb{R}^d} \frac{1}{|D|} \sum_{\zv \in D} l(\xv; \zv),
\end{equation}
where $\xv \in\mathbb{R}^d$ represents the ML model variable that we aim to optimize. The loss function $l(\xv; \zv)$ is evaluated at model $\xv$ and data sample $\zv=(\vect{z}_{\text{in}}, z_{\text{out}})$, which describes the input-output relationship between $\vect{z}_{\text{in}}$ and its label $z_{\text{out}}$, and $F: \mathbb{R}^d\rightarrow\mathbb{R}$ denotes the differentiable loss function averaged over the entire dataset $D$. It is assumed that there exists a latent distribution $\nu$ that governs the generation of the global dataset $D$, where each data sample $z \in D$ is independently and identically distributed (IID)\footnote{In  Section~\ref{sec:sim} we will numerically evaluate non-IID datasets.} from $\nu$.  We denote
$
    \xv^* \triangleq \argmin_{\xv \in \mathbb{R}^d} F(\xv), f^* \triangleq F(\xv^*).
$

One category of distributed and decentralized ML, including FL, aims to solve the ERM problem \eqref{eqn:erm} by utilizing a set of clients that perform local computations in parallel, leading to improved wall-clock speed compared to the centralized training paradigm. In our distributed ML system, we consider a central parameter server (e.g., at the base station) and a set of $n$ clients (e.g., IoT devices). Mathematically, problem \eqref{eqn:erm} can be equivalently expressed as
 \begin{equation} \label{eqn:erm2}
\min_{\xv \in \mathbb{R}^d} F(\xv) = \min_{\xv \in \mathbb{R}^d} \frac{1}{N} \sum_{i=0}^{N-1}  F_i(\xv),
\end{equation}
where $F_i(\xv)$ represents the local loss function at client $i$, defined as the average loss over its respective local dataset $D_i$: $F_i(\xv) = \frac{1}{|D_i|} \sum_{\zv \in D_i} l(\xv; \zv)$. We assume that the local datasets are disjoint, and their union results in the global dataset $D = \cup_{i \in [n]} D_i$. In this work, we primarily focus on the \emph{full clients participation} setting, where all $N$ clients participate in every round of distributed SGD. However, we will report numerical results for partial client participation in Section~\ref{sec:sim}. For simplicity and ease of analysis, we also assume that all clients have equal-sized local
datasets, i.e., $|D_i|=|D_j|, \forall i, j \in [n]$.

\subsection{The \textsl{FedAvg} Pipeline}
\label{sec:modelFedAvg}

\begin{figure}
    \centering
    \includegraphics[width=1\linewidth]{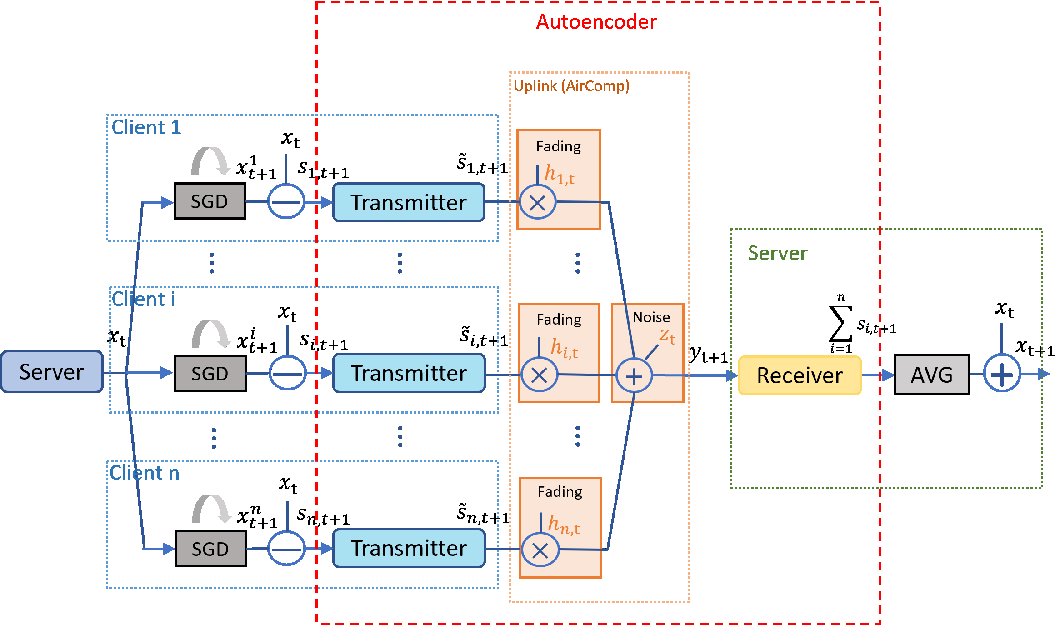}
    \caption{\small Federated learning pipeline in the $t$-th global round.}
    \label{fig:SystemDiagram}
    \vspace{-0.2in}
\end{figure}

The distributed ML paradigm we consider follows the foundational framework of \textsl{FedAvg} \cite{mcmahan2017fl}, with a specific focus on incorporating fading channels in the upload communication phase. The system diagram depicting the overall architecture is presented in Fig.~\ref{fig:SystemDiagram}.  In particular, the \textsl{FedAvg} pipeline operates by iteratively executing the following steps during the $t$-th learning round, where $t$ belongs to the set $[T] \triangleq \sth{1, 2, \cdots, T}$. 


\mypara{(1) Download the global model.} The server broadcasts the current global model $\xv_t$ to all clients. In the context of a wireless FL setting, the base station typically possesses more transmit power and communication resources compared to the clients, which are mostly battery-powered mobile devices with limited capabilities. As a result, it is commonly assumed that the download communication phase is error-free \cite{zhu2019broadband,yang2020federated,amiri2020sp,amiri2020federated2,sery2020tsp}. Following this assumption, all clients receive an accurate copy of $\xv_t$, ensuring that they possess the exact knowledge of the global model.

\mypara{(2) Local model update at clients.} Each client $i$ updates the received global model $\xv_t$ based on its local dataset $\Dc_i$, assumimg the use of SGD for model training. 

Specifically, SGD at client $i$ proceeds by updating the weight iteratively (for $E$ steps in each learning round) as follows:
{\small
    \begin{align*}
        \text{Initialization: }&  \vect{x}_{t,0}^i = \vect{x}_{t}, \\
        \text{Iteration: }&  \vect{x}_{t,\tau}^i = \vect{x}_{t,\tau-1}^i - \eta_t \nabla  f_i(\vect{x}_{{t},\tau-1}^i), \forall \tau=1, \cdots, E,\\
        \text{Output: }&  \vect{x}_{t+1}^i =\vect{x}_{{t},E}^i,
    \end{align*}
}%
\vspace{-0.1in}
where we define
\begin{equation} \label{eqn:ffi}
f_i\pth{\xv} \triangleq  l \pth{\xv; \xi_i}, \qquad    f\pth{\xv} \triangleq  l \pth{\xv; \xi}.
\end{equation}
Here $\xi_i$ and $\xi$ represent data points sampled independently and uniformly at random (u.a.r.) from the local dataset of client $i$ and the global dataset, respectively. It is worth noting that this formulation can be easily extended to \emph{mini-batch SGD} by allowing $\xi_i$ and $\xi$ to represent a batch of data points sampled u.a.r. from $\Dc_i$ and $\Dc$ without replacement, respectively.

\mypara{(3) Upload local models.} After local model update, client $i$ needs to calculate the model differential parameter
\begin{equation} 
\label{eqn: ulsymb}
s_{i,t+1} \triangleq \vect{x}_{t+1}^i - \vect{x}_{t}
\end{equation}
 as the input of each transmitter. And then the $s_{i,t+1}$ is modulated through $f_i$ to generate the transmitted signal $\tilde{\vect{s}}_{i,t+1}\in\mathbb{R}^m$. Then all the transmitter signals are sent synchronously over the uplink wireless channel to the parameter server.
Finally, the receiver applies the transformation $g$ to recover the sum of the model differential parameters $\sum_{i=1}^{n} s_{i,t+1}$. The fading channel model and the transmitter/receiver processing are described later in Section~\ref{sec:uplink_comm} and we re-emphasize that this paper focuses on \emph{digital} model communications through AirComp.

\mypara{(4) Global aggregation.}  The server aggregates the received local models summation $\sum_{i=1}^{n} s_{i,t+1}$ to generate a new global ML model $\xv_{t+1}$.  
\begin{equation*}
\xv_{t+1} = \xv_{t} + \frac{1}{n} \sum_{i=1}^{n} s_{i,t+1}
\end{equation*}
for any $t \in [T]$.
For simplicity, we assume that each local dataset has an equal size as the original design of \textsl{FedAvg} \cite{mcmahan2017fl}, which means choosing the equal weight for every client. It is conceivable to extend the proposed method to unequal weights by processing the dataset size info at the uploading phase.

\subsection{Communication over Fading Channels}
\label{sec:uplink_comm}
\vspace{0.5em}
We now elaborate on the upload communication phase in the previous section, where each client $i$ aims at sending vector $\tilde{\vect{s}}_{i,t+1}$ to the server over wireless fading channels. The illustration of power control will be discussed in Section~\ref{subsec:autoencoder}. We consider that the transmission of $\tilde{\vect{s}}_{i,t+1} $ experiences a random fading channel fluctuation $h_{i,t}$ for each of its $d$ elements. This assumption is valid when the underlying channel follows a \emph{block fading} model \cite{Goldsmith:05}. In this model, the channel remains constant for a duration of at least $d$ symbol periods, ensuring that the coherence time is longer than $d$. After this duration, the channel changes independently to another value following its distribution, such as a Gaussian distribution.

Then all the signals are sent to the parameter server in a superpositioned manner through AirComp, and we finally have the received signal
\begin{equation} \label{eqn:ulrxsymb}
\tilde{\vect{y}}_{t+1} = \sum_{i=1}^{n} h_{i,t} \tilde{\vect{s}}_{i,t+1} +  \vect{z}_t
\end{equation}
where $\vect{z}_t$ denotes an additive white Gaussian noise (AWGN) vector with $d$ independent Gaussian elements of mean zero and variance $N_0$. 

We assume each individual client has perfect channel state information at the transmitter (CSIT) and a {\em coherent receiver} with perfect channel state information at the receiver (CSIR). Hence, the client can eliminate the effects of channel fading by scaling the transmitted signal $\tilde{\vect{s}}_{i,t+1}/h_{i,t} $ before experiencing the fading channel. Subsequently, the receiver computes an unbiased estimate of the summation of $\vect{s}_{i,t+1}$ as
\begin{equation} \label{eqn:ulrxsymb2}
\vect{y}_{t+1} = \sum_{i=1}^{n} h_{i,t} (\tilde{\vect{s}}_{i,t+1}/h_{i,t})  +  \vect{z}_t = \sum_{i=1}^{n}\tilde{\vect{s}}_{i,t+1} + \vect{z}_t
\end{equation}

The goal of this paper is to design an end-to-end communication system by an autoencoder that can restore the summation of model differential parameters, which is emphasized with the red-dotted rectangle in Fig.~\ref{fig:SystemDiagram}.

\subsection{Problems with Existing Constellations}
\label{sec:pro_const}
\vspace{0.5em}

In modern communication systems, the input signal is digitized to enable efficient transmission. Information bits are encoded and then modulated. Upon modulation, each symbol is represented by a complex number (chosen out of a finite number of candidates), denoting a specific point in the constellation space. The real and imaginary parts of the complex number are mapped to the horizontal and vertical axes, respectively. 
When multiple clients are involved in AirComp, each client will have their individual constellation (e.g., Quadrature Phase Shift Keying (QPSK)), and the receiver ideally gets the sum of the constellations from all involved clients.  Unlike analog signals where the sum can be represented as a real number, special attention must be given to the signal constellation design to ensure accurate decoding of the sum. It is crucial to avoid situations where a particular point in the (sum) constellation corresponds to different sums.

\begin{figure}
    \centering
    \includegraphics[width=0.26\textwidth]{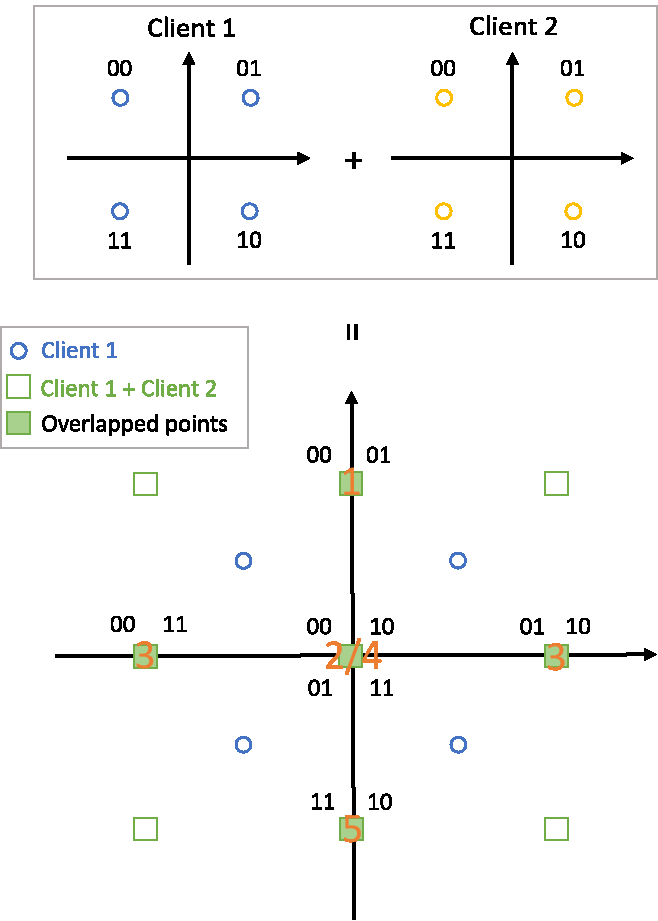}
    \caption{\small An example of overlapped constellation points.}
    \label{fig:overlap}
    \vspace{-0.25in}
\end{figure}

In Fig.~\ref{fig:overlap}, we highlight an example of a signal constellation for a QPSK modulation. In this example, there are 16 candidates for decoding (4 bits per symbol), and 12 of them are overlapped with the other 5 constellation points, indicated by the green markers. While most of the overlapped points retain a unique sum of the two signals, ambiguity arises for the midpoint, which can correspond to either 2 or 4, making it impossible to determine the exact sum. We note that such ambiguity would become more severe with higher-order modulation and more clients, as the possible combination grows exponentially.

To address this challenge, one approach is to manually design the constellation to avoid ambiguity. However, as the number of clients and the order of modulations increase, the computational complexity of handcrafted designs quickly becomes impractical. Thus, we explore the potential of employing autoencoder techniques to automatically design the required constellations. This will be elaborated in the next section.

\section{End-to-End Communication System}
\label{sec:end_to_end}
The key idea of the autoencoder-based communication systems is to unify the transmitters, channel, and receiver via jointly training the autoencoder network. Unlike traditional autoencoders proposed for communication systems, our autoencoder aims to reproduce the total sum of its input as the output. In its most basic configuration, the communication system comprises $n$ transmitters, a channel, and a receiver, as illustrated in Fig.~\ref{fig:AutoencoderDiagram}. 

Let the size of the model differential $\vect{s}_{i,t+1}$ be $d$, and we reshape the
parameters $\{\vect{s}_{i,t+1}\}_{i=1}^n$ to vectors $\{\vect{s}_{i,t+1} \in \mathbb{R}^d \}_{i=1}^n $ for each transmitter. In order to transmit the d-dimensional source message to the server in the FL setting, a series of processing steps are employed. First, we quantize the source vector into discrete values, and then apply source coding and modulation independently in each client. Once the server receives the summation of transmitted signals from all the clients through the over-the-air computation, demodulation is performed to recover the sum of the quantized message, and dequantization is applied to retrieve the original signal. The modulation and demodulation operations are respectively implemented by the encoders and the decoder utilized in our approach, as shown in Fig.~\ref{fig:AutoencoderDiagram}. 
\subsection{Quantization and Encoding}
\label{subsec:quant}
In ML models such as deep neural networks (DNN), the model weights are typically represented in a 32-bit floating point format. However, in FL settings, communication between devices can become a bottleneck due to the large size of the transmitted messages. To address this issue, quantization techniques can be applied to reduce the bit-width required for each weight and, as a result, decrease the message size for communication. The appropriate quantization method is important but not the focus of this paper. Instead, we adopt the effective FL quantization method in \cite{zheng2020design}. Specifically, for a full-precision weight $\vect{s}_{i,t+1}$, a $k$-bits quantization is completed via the following steps (to simplify the notation, we use $\xv^i$ to represent $\vect{s}_{i,t+1}$):

\mypara{(1) Scale Up.}Each element of $\xv^i$ is first amplified by a scaling factor known as the quantization gain $G$. This yields the amplified value $\xv_a^i = \xv^i*G$. Typically, $G$ is
set as power of 2, simplifying the implementation through bit shifting.

\mypara{(2) Stochastic Rounding.}The amplified value $\xv_a^i$ is then rounded to its integer part using stochastic rounding. The rounding function $R(\cdot)$ selects either the upper bound $\lceil\xv_a^i\rceil$ or the lower bound $\lfloor\xv_a^i\rfloor$ with certain probabilities. 
More formally, we have ($w.p.$ is short for `with probability):
\begin{equation}
R(\xv_a^i)=
\begin{cases}
  \lfloor\xv_a^i\rfloor, & \text{w.p. } 1-(\xv_a^i-\lfloor\xv_a^i\rfloor)\\
   \lceil\xv_a^i\rceil, &  \text{w.p. } \xv_a^i-\lfloor\xv_a^i\rfloor \\
\end{cases}
\end{equation}

\mypara{(3) Limit.}The range of the rounded integer value $\xv_r^i$ is further restricted to $k$ bits. Specifically, we have:
\begin{equation}
\xv_l^i=
\begin{cases}
    -2^{k-1}, & \text{if } \xv_r^i < -2^{k-1}\\
   \xv_r^i, &   \text{if } \xv_r^i \in [ -2^{k-1}, 2^{k-1}-1]\\
   2^{k-1}-1, & \text{if } \xv_r^i>2^{k-1}-1\\
\end{cases}
\end{equation} 
For example, if $k=2$ bits, the quantized $\xv_l^i \in [-2,-1,0,1]$.

\mypara{(4) Scale Down.}The receiver obtains the output $\xv_{output}$ by scaling
down the input $\xv_{input}: \xv_{output} = \xv_{input}/G$. Note that this step is the dequantization process in the server.

To ensure that the quantized signal with $k$ bits is compatible with the autoencoder input during encoding, the quantized signal $\xv_l^i$ is adjusted to start from 0 by adding a constant number $2^{k-1}$ to it. This adjustment results in $\xv_l^i$ becomes to $\xv_q^i \in [0,1,2,3]$ when $k$ is 2 bits.

\subsection{Autoencoder}
\label{subsec:autoencoder}

\begin{figure*}
    \centering
    \includegraphics[width=0.58\textwidth]{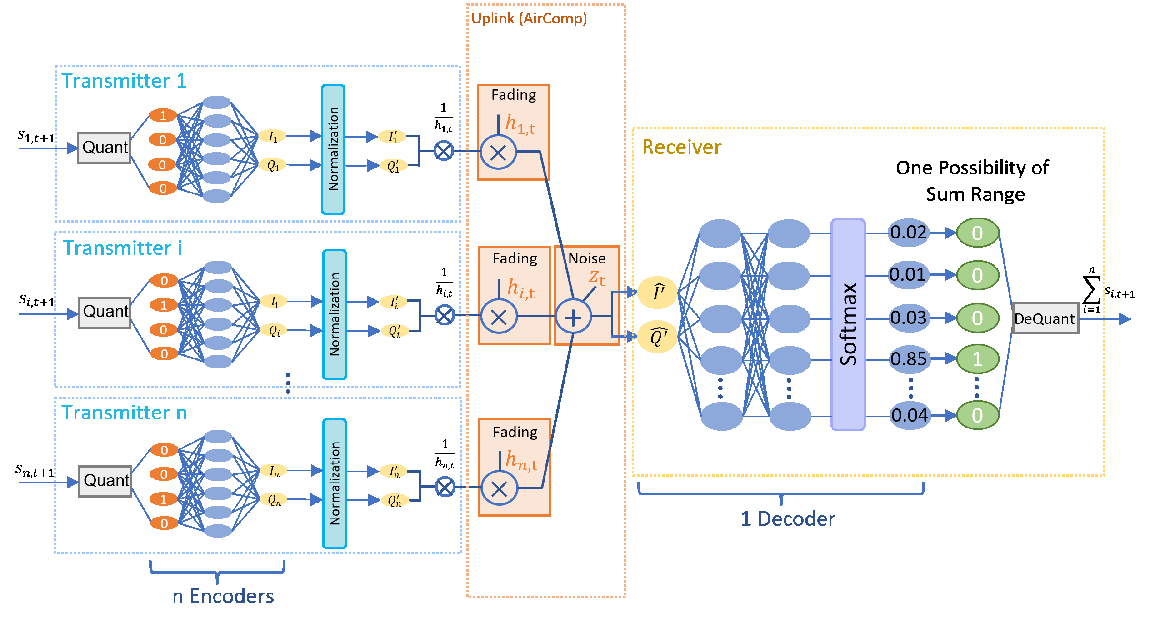}
    \caption{\small End-to-end communication system realized with autoencoder.}
    \label{fig:AutoencoderDiagram}
    \vspace{-0.2in}
\end{figure*}

Since the number of bits per message $\xv_q^i$ in the transmitter $i$ is denoted as $k$, each transmitter aims to convey a single message from a set of $M=2^k$ feasible messages, $\xv_q^i \in \mathbb{M} = \{0, 1, ..., M-1\} \forall i$, to the receiver through $\mv$ distinct uses of the communication channel. In this paper, we use one-hot encoded vectors as input to the encoder, which is an M-dimensional vector with a single element being one and the other elements being zero.

A complex signal consists of two real signals - one for the real part $I$ and one for the imaginary part $Q$. To simplify the analysis, we focus on the real-valued signals only. Consequently, we map the complex domain of $\mathbb{C}^{\frac{1}{2}m}$ to the $\mv$-dimensional Euclidean space of $\mathbb{R}^m$. Therefore, the transmitter applies $f_i: \mathbb{M} \rightarrow \mathbb{R}^m$ to the message $x_q^i$ to generate the transmitted signal $\sv^i=f(\xv_q^i) \in\mathbb{R}^m$. 

In typical communication systems, the hardware of the transmitter applies power constraints to the transmitted signal to maintain the system's reliability and efficiency. Brianna et al. discuss the impact of both hard and soft power constraints on system performance \cite{robertson2021optimizing}. In our work, we focus on the hard constraint, whereby the transmitter is limited to a maximum energy threshold of $P$ for all signal constellation points, i.e., $|s_j^i| \leq m ~\forall j$. Hence, the normalized signal 
\begin{equation} \label{eqn:normalized}
\tilde{\vect{s}_j^i} = \sqrt{\frac{1}{\Expect \norm{\vect{s}_j^i}^2}} P\vect{s}_j^i= \sqrt{\frac{1}{\Expect \norm{\vect{s}_j^i}^2}} m\vect{s}_j^i.
\end{equation}

The corresponding received SNR for the signal $i$ is
\begin{equation} \label{eqn:ulrxsnr}
\ssf{SNR}_{t} = \frac{m}{N_0}
\end{equation}
where $N_0$ is the variance of Gaussian noise $\zv_t$ in \eqref{eqn:ulrxsymb}.

The communication rate of this communication system is $R = k/m$ bits per real channel use, where $k = log_2M$. Traditionally, the notation $(m,k)$ means that each transmitter sends one out of $M = 2^k$ messages (i.e., k bits) through $\mv$ real channel uses to the receiver. The normalized modulated signals presented with $I_i^{'}$ and $Q_i^{'}$ in Fig. \ref{fig:AutoencoderDiagram} pass through the individual fading channel and are added together. The channel is described in Section \ref{sec:uplink_comm}, where $\vect{y}_{t+1} \in \mathbb{R}^m$ denotes the received signal. 

Upon reception of $\vect{y}_{t+1}$, the receiver applies the transformation $g: \mathbb{R}^m \rightarrow \mathbb{S}$ to produce the estimate of the sum of the transmitted message $\sum_{i=1}^{n} \tilde{\vect{s}}_{i,t+1}$. Let $\mathbb{S}$ denote the set of all feasible sums of messages $x_q^i$ from a set of $n$ transmitters in a communication system. Consider, for instance, a communication system with $n=8$ transmitters that sends a $k=2$ bits message. In this scenario, the exact set of possible values for the summation is given by $\mathbb{S}=\{0, 1, ..., 8(M-1)\}$, where $M=2^k=4$. The cardinality of $\mathbb{S}$ is 25, which represents the total number of possible summations in this communication system. Therefore, we set the last layer of $g$ as a softmax layer to ensure that the output activations form a probability vector over $M$. Finally, the index with the highest probability is chosen as one of the possibilities of the set $\mathbb{S}$. A generic architecture of the decoder is shown in Fig. \ref{fig:AutoencoderDiagram}.

\subsection{Advanced Design for Higher-Order Modulations}
\label{subsec:more_bits}
In Section \ref{subsec:autoencoder}, we present our label set $\mathbb{S}$  and define the function representing the total possible sum. The number of classes is determined by $n(2^k-1)+1$, where n denotes the number of clients. Consequently, as the number of bits increases, the categories to be classified experience exponential growth. This exponential increase in the number of classes poses a significant challenge in accurately predicting a specific class, particularly with higher-order modulations. To address this challenge, we propose streamlining categorization, aiming to enhance tolerance to errors by reducing the number of classes. The subsequent paragraphs will provide a detailed description of both methods. To facilitate our analysis, we set $n=8$ clients and $k=4$ bits.

The motivation behind streamlining categorization is to ensure the scalability of our autoencoders. By maintaining a consistent number of classes, irrespective of the number of bits per message, we can establish a standardized framework. For instance, we can consider 25 classes equivalent to $k=2$ bits. Recognizing the difficulty in accurately predicting all $121$ possible 4-bit sums, we adopt a coarse-grained classification approach. Specifically, we divide these $121$ types into an average of 25 categories, with each category representing a range of approximately five numbers centered around the median value. While label errors exist, their impact on performance is mitigated by quantization error, SGD noise, and channel noise. Consequently, this method remains effective even in adverse channel conditions characterized by a low SNR, a finding supported by our experimental validation.

\section{Experimental Results}
\label{sec:sim}
\subsection{Setup}
\label{sec:settup}

We have carried out experiments to evaluate our method on the popular dataset: 
CIFAR-10 \cite{krizhevsky2009learning} (60,000 images with 10 classes). We report experimental results for both IID datasets and non-IID datasets with full client participation. In the experiments, our autoencoder training is conducted at a fixed SNR value of 7 dB (refer to Section \ref{subsec:autoencoder}), utilizing the Adam \cite{kingma2014adam} optimization algorithm with a learning rate of 0.001. And for the FL setting, we set m=8 clients.

We consider the following schemes in the experiments. (1) \textbf{Perfect\_Comm}: the {ideal} case with perfect communication. (2)~\textbf{AE\_opt SNR}: our proposed method with different SNRs. All of the reported results are
obtained by averaging over five independent runs

\subsection{2-bit Modulation}
\begin{figure}[htbp]
\vspace{-0.1in}
\centering
\includegraphics[width=0.58\linewidth]{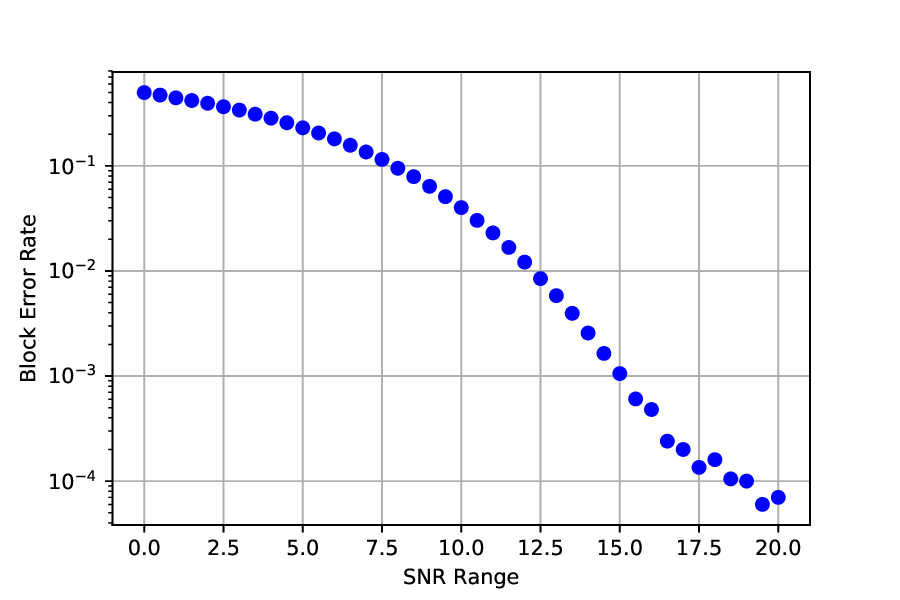}\hfill
\includegraphics[width=0.39\linewidth]{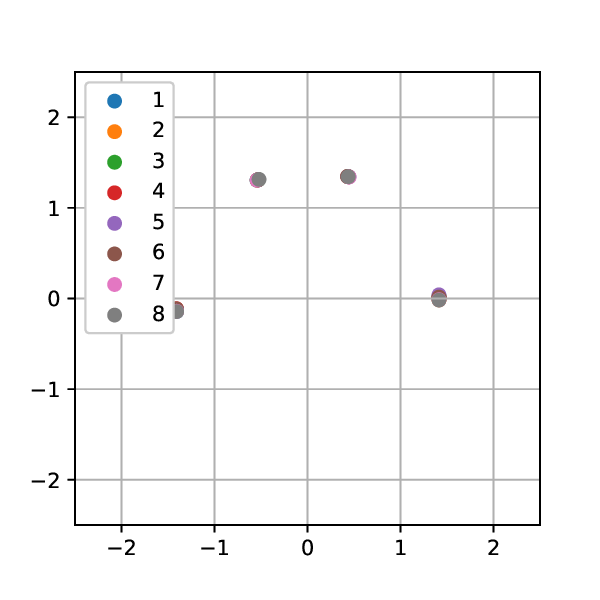}
\hspace*{25pt} \footnotesize (a) BLER vs. SNR \hspace{50pt}  (b) Constellation diagram
\caption{\small  Autoencoder results using parameters (2,2) with 8 clients.}
\label{fig:8clients_22}
\vspace{-0.1in}
\end{figure}

Fig.~\ref{fig:8clients_22}(a) shows our (2,2) autoencoder block error rate (BLER), i.e., Top-1 classification error, versus the SNR. And Fig.~\ref{fig:8clients_22}(b) shows the learned representations $\tilde{\vect{s}}$ of all messages from 8 clients for the parameter of (2, 2) as complex constellation points, i.e., the x- and y-axes correspond to the first and second transmitted symbols, respectively. Notably, the constellation designs for different clients exhibit significant similarity, thus facilitating the success of random transmitter selection when a subset of clients participate in the FL training round.

\begin{figure}[htbp]
\centering
\includegraphics[width=0.49\linewidth]{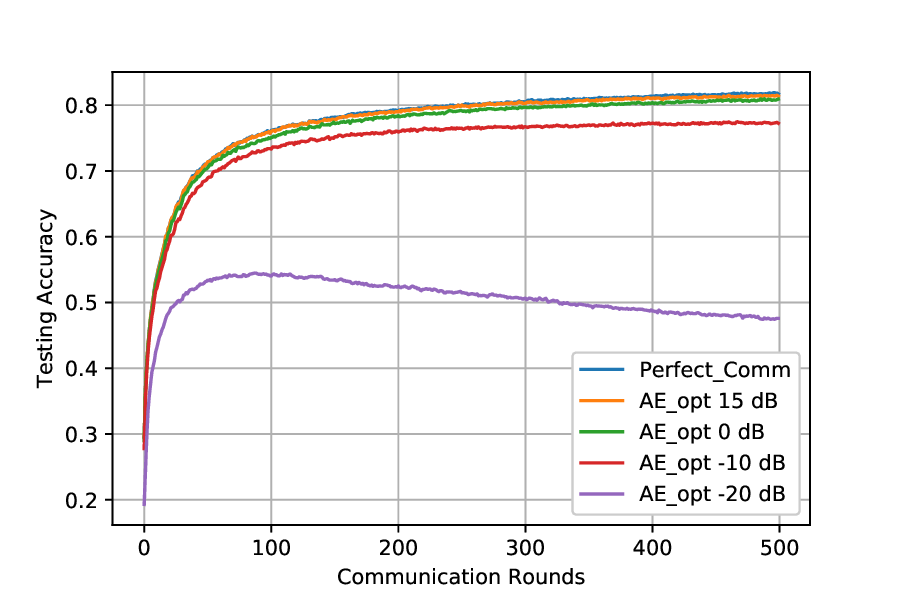}\hfill
\includegraphics[width=0.49\linewidth]{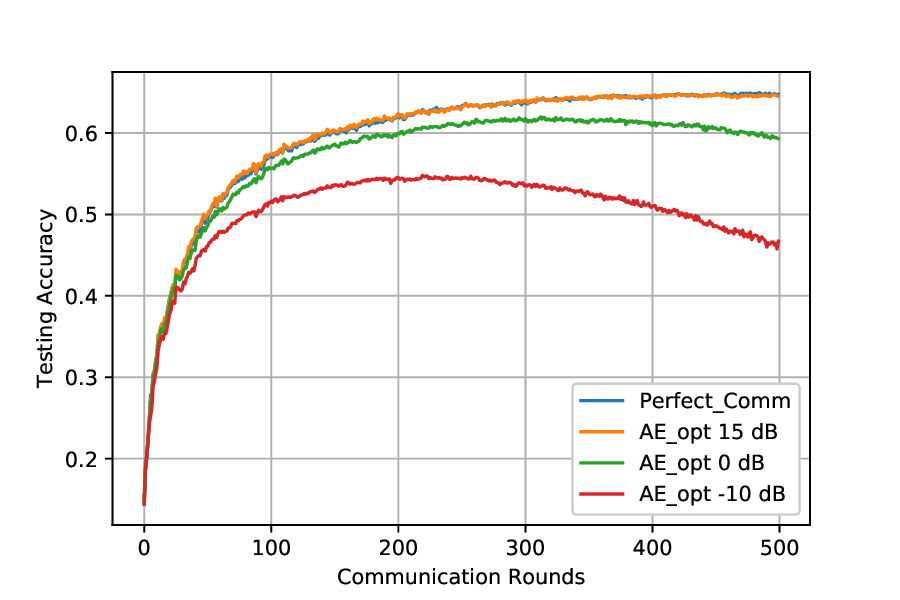}
\hspace*{2pt} \footnotesize (a) IID \hspace{95pt}  (b) Non-IID
\caption{\small 2-bit modulation of CIFAR-10 results with IID and Non-IID local datasets and full clients (8 clients) participation.}
\label{fig:8clients_22_FL}
\vspace{-0.1in}
\end{figure}

Subsequently, we apply the (2,2) autoencoder to the FL setting. As depicted in Fig.~\ref{fig:8clients_22_FL}(a), the proposed AE\_opt achieves close performance to Perfect\_Comm at SNR values of 15 dB and even 0 dB, demonstrating the effectiveness of our autoencoder. Even under extremely adverse conditions, such as an SNR of -10 dB, the top-1 accuracy remains relatively high, experiencing only a minor drop compared to Perfect\_Comm. Additionally, when considering Non-IID local datasets in Fig.~\ref{fig:8clients_22_FL}(b), our AE\_opt with an SNR of 15 dB maintains comparable performance to Perfect\_Comm. In addition, we observe a decrease in accuracy as the number of training rounds increases in the non-IID scenario. This accuracy drop occurs because the range of model differentials becomes smaller when the model is close to convergence. Therefore, careful consideration must be given to the choice of quantization gain. Interestingly, we find that using a larger quantization gain in the later rounds improves performance, resulting in better accuracy. This observation suggests that adjusting the quantization gain during the training process can help mitigate the accuracy degradation associated with non-IID scenarios.
\vspace{-0.1in}

\subsection{Higher-order Modulations}

\begin{figure}[htbp]
\centering
\includegraphics[width=0.49\linewidth]{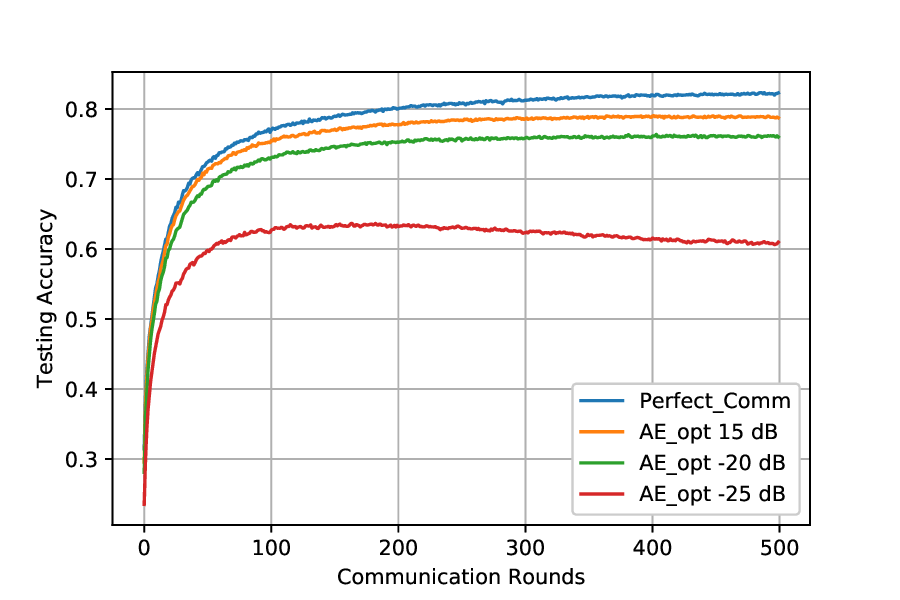}\hfill
\includegraphics[width=0.49\linewidth]{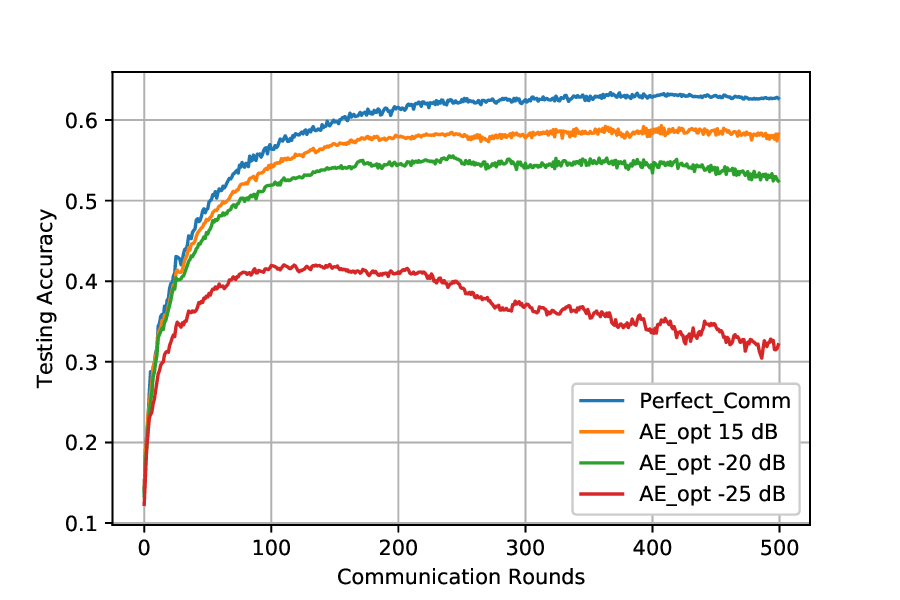}
\hspace*{2pt} \footnotesize (a) IID \hspace{95pt}  (b) Non-IID
\caption{\small 4-bit modulation of CIFAR-10 results with IID and Non-IID local datasets and full clients (8 clients) participation.}
\label{fig:8clients_42_25types_FL}
\vspace{-0.1in}
\end{figure}

We conduct further validation of our advanced autoencoder design specifically tailored for higher-order modulations, such as 4 bits. To begin, we employ our streamlining categorization method to train a (4,2) autoencoder. 
It is observed in Fig.~\ref{fig:8clients_42_25types_FL} that AE\_opt at an SNR of 15 dB demonstrates convergence and achieves comparable performance for both IID and Non-IID local datasets. These results provide empirical evidence to support our previous hypothesis that the presence of labeling errors can be mitigated by factors such as quantization error, SGD noise, and channel noise, thus minimizing their impact on FL performance. Remarkably, this method remains effective even under challenging channel conditions characterized by a low SNR.
\vspace{-0.1in}

\section{Conclusion}
\label{sec:conc}
We have proposed a novel end-to-end communication system that utilizes AirComp for digital transmission, with the objective of accurately decoding the sum within the constellation design. Through the incorporation of autoencoder network structures and the joint optimization of transmitter and receiver components, our approach effectively addresses the challenges associated with deploying AirComp techniques in the context of digital modulation-based wireless communications. The implications of our research hold significant importance for the field of wireless FL, as we establish a robust foundation for future investigations and provide practical insights into enhancing the efficiency and scalability of FL in wireless environments. Our experimental results demonstrate that our AE\_opt achieves near-perfect communication performance in high SNR scenarios. Furthermore, the performance of AE\_opt can be further improved by refining the autoencoder design for higher-order modulations.

\bibliographystyle{IEEEtran}
\bibliography{refs/reference}

\end{document}